\documentclass{jpsj2}
\title{
Fluctuation Effect in the $\pi$-Flux State for Undoped High-Temperature Superconductors
}

\author{Takao Morinari
\thanks{E-mail: morinari@yukawa.kyoto-u.ac.jp}}

\inst{Yukawa Institute for Theoretical Physics, Kyoto University, Kyoto 606-8502, Japan}

\abst{
The effect of fluctuations about the $\pi$-flux mean field state 
for the undoped high-temperature superconductors is investigated.
It is shown that fluctuations of the mean fields lead to a self-energy correction 
that doubles the band width of the fermion dispersion in the lowest order.
The dynamical mass generation is associated with the self-energy effect due to
the interaction mediated by the Lagrange multiplier field,
which is introduced to impose the constraint on the fermions.
A self-consistent picture about the mass generation and the propagation of the 
Lagrange multiplier field without damping is proposed.
The antiferromagnetic long-range ordering is described 
without introducing an additional repulsive interaction.
The theory suggests a natural framework to study spin disordered systems
in which fermionic excitations are low-lying excitations.
}

\kword{high-temperature superconductivity, $\pi$-flux state, 
antiferromagnetic Heisenberg model, dynamical mass generation,
confining potential
}

\begin{document}
\maketitle

\newcommand{\be}{\begin{equation}}
\newcommand{\ee}{\end{equation}}
\newcommand{\bea}{\begin{eqnarray}}
\newcommand{\eea}{\end{eqnarray}}

\section{Introduction}
In the phase diagram of the high-temperature superconductors, 
apparently the most established phases are 
the $d$-wave superconducting phase and the N{\' e}el ordering phase
in the undoped compound.
One of the key questions about high-temperature superconductivity
is how to connect these two phases.
If one starts from the N{\' e}el ordering state toward the d-wave superconducting
state, the first step would be to consider the single hole doped system.


Experimentally such the system has been studied by
angle-resolved photoemission spectroscopy (ARPES) in the undoped compounds
\cite{Wells1995,LaRosa1997,Ronning1998}
where a photo-hole is introduced in the system and the excitation spectrum 
is associated with the properties of the single hole doped system.
The experimentally obtained excitation spectrum is 
in qualitatively good agreement
with the quasiparticle excitation spectrum of the $\pi$-flux mean field state 
\cite{AffleckMarston1988,Affleck1989} with a mass term.
\cite{Hsu1990}
Although at the mean field level the quasiparticles are gapless,
an excitation energy gap opens up by adding an on-site repulsive
interaction.
The gap arises from the staggered magnetization,
and by taking it as a variational parameter,
better variational energy is obtained.\cite{Hsu1990}
An effective field theory approach also suggests
the presence of the mass term.\cite{KimLee1999}
In addition, the dispersions along $(0,0)$-$(\pi,\pi)$ line
and $(0,\pi)$-$(\pi,0)$ line in the Brillouin zone are 
isotropic as observed
in the experiments.\cite{Ronning1998}
(The lattice constant is taken as the unit of length throughout the paper.)
However, the band width of the quasiparticles in the $\pi$-flux mean field
state is smaller than the experiment by a factor of $2-3$.
Furthermore, there is no reliable estimation of the mass value 
except for variational wave function approaches.
However, in the variational wave function approaches it is 
necessary to include an additional repulsive interaction.

On the other hand, self-consistent Born approximation analysis of the t-J model
\cite{KaneLeeRead1989}
suggests that the band dispersion is significantly renormalized from 
$t \simeq 0.4$eV to $J \simeq 0.13$eV, which is consistent with
the experiments.
However, there is some discrepancy in the excitation spectrum
along the line in the momentum space from $(\pi,0)$ to
$(0,\pi)$. 
This discrepancy is removed by including the next-nearest and 
the third nearest neighbor hopping processes.\cite{TohyamaMaekawa2000}
In this approach, the t-J model is analyzed in terms of the 
slave-fermion mean field theory.
In the slave-fermion mean field theory, the spins are described by
the Schwinger bosons.\cite{Arovas1988}
Therefore, it is straightforward to describe the N{\' e}el ordering state
as Bose-Einstein condensate of those bosons.
However, to describe the d-wave superconducting state the slave-fermion formalism
is not convenient. 
For the description of the d-wave superconducting state, slave-boson formalism
is used in the literature.\cite{LeeNagaosaWen2006}
In order to avoid taking a different formalism, here I focus on the $\pi$-flux state.

Before going into discussions about the $\pi$-flux state,
let us discuss advantages and disadvantages of the theory.
In order to consider the single hole starting from the spin $1/2$ 
antiferromagnetic Heisenberg model, mainly there are two approaches:
One is to represent spins in terms of boson fields,
like in the non-linear $\sigma$ model \cite{Chakravarty1988}
or equivalently in the CP$^1$ model \cite{Read1990} 
derived from the Schwinger boson mean field theory.\cite{Arovas1988}
In this approach, we need to think of a soliton-like excitation
to describe a doped hole, which is fermion, in terms of
boson fields.
The situation is very similar to the Skyrme model \cite{Skyrme1962}
in which fermions (protons and neutrons) are described as 
solitons in the non-linear $\sigma$ model.
(Goldstone modes, which correspond to spin wave excitations in the
Heisenberg antiferromagnet, are pion fields.)
It is argued that in ref. \citenum{Morinari2005} that a doped hole
can be described by a skyrmion-like spin texture.
\cite{Gooding1991,Waldner1986,Belov1998,Haas1996,Timm2000,Marino2000,Moskvin1999,Seibold1998,Pereira2007,Baskaran2003}
The band width and the mass, which is the excitation energy
of the spin texture,
are in good agreement with experimentally estimated values.
(In the experiments, 
the mass value can be estimated by approximating the dispersion
around $(\pm \pi/2, \pm \pi/2)$ by a conventional non-relativistic 
kinetic energy form.)
From the Ginzburg-Landau description of the Heisenberg antiferromagnet,
it is natural to expect the appearance of such a spin texture
in the continuum:
In the bosonic theory
the N{\' e}el ordering state is described by 
a Bose-Einstein condensate.
It turns out that local suppression of the order parameter,
which is introduced by the formation of the Zhang-Rice singlet,\cite{ZhangRice1988}
leads to a vortex-like state.
This solution is easily found from the analysis of two-dimensional
Ginzburg-Landau theory.\cite{FetterWalecka}
From the CP$^1$ formalism,\cite{Rajaraman} 
the vortex like state turns out to be a skyrmion
like spin texture.
However, stability of such a state on the lattice is not evident.
This is because that the bosonic theory describes mainly 
the low-lying excitations,
that is, the spin wave excitations.
Therefore, it is not easy to study microscopic stability 
of the spin texture within a bosonic theory.

By contrast, the other approach is to describe spins in terms
of fermions:
\be
S_{j\alpha} = \frac{1}{2} f_j^{\dagger} \sigma_{\alpha} f_j,
\label{eq_spin_rep}
\ee
where 
$f_j^{\dagger}=\left(f_{j\uparrow}^{\dagger}, f_{j\downarrow}^{\dagger}\right)$,
$\sigma_{\alpha}$ ($\alpha=x,y,z$) are the Pauli matrices.
Because the spins are $1/2$, there is a constraint on the fermions,
\be
\sum_{\sigma=\uparrow,\downarrow} f_{j\sigma}^{\dagger} f_{j\sigma}=1.
\label{eq_constraint}
\ee
It is expected that the quantum nature of the spin $1/2$
is well described by this theory.
However, there is disadvantage in the description
of the antiferromagnetic long-range ordering.
Within the mean field theory, the staggered magnetization vanishes.
Therefore, we need to go beyond the mean field theory
to describe N{\' e}el ordering.
In the continuum limit, the N{\' e}el ordering is discussed \cite{KimLee1999}
in the context of the dynamical mass generation in quantum electrodynamics 
in three spatial and time dimensions.
\cite{Pisarski1984,Appelquist1986,Nash1989,Appelquist1988,
Gusynin2003}
Although there are discussions about the condition of the mass generation 
in the sense of the 1/N expansion, 
it is hard to estimate the value of the dynamically generated mass.
Another way to describe the mass generation is to include
the short-range Coulomb repulsion.\cite{Hsu1990}
However, it is not clear whether adding the short-range Coulomb repulsion term 
to the antiferromagnetic Heisenberg model is necessary.
The antiferromagnetic Heisenberg model is derived from the Hubbard model
taking the limit in which the on-site Coulomb repulsion, $U$,
is much larger than the hopping matrix element, $t$.
The condition of $U/t \gg 1$, is replaced by the constraint.
So, it is unclear whether an additional repulsive interaction is necessary.

In this paper, we study the fluctuation effects in the $\pi$-flux state.
The constraint is included in the action of the system using a standard 
Lagrange multiplier field.
The effect of the constraint is studied by analyzing the self-energy effect
associated with fluctuations of the Lagrange multiplier fields.
It is argued that by including fluctuations the resulting quasiparticle
band width and the staggered magnetization are in good agreement
with the experiment.
The result suggests that the $\pi$-flux state is a promising candidate
for the description of the undoped high-temperature superconductors.
In particular, the fact that the model described by the fermion fields
includes the N{\' e}el ordering state
implies that the theory provides us a natural framework to study
doping effect and the spin disordered phase in the presence of the 
doped holes.
Because the quantum nature of the holes are well described by 
such the theory.

The organization of this paper is as follows.
In Sec.\ref{sec_formalism}
we study the fluctuations of the $\pi$-flux state mean fields.
It is shown that the lowest order self-energy correction 
doubles the amplitude of the mean fields.
In Sec.\ref{sec_cpi}
the coherent state path-integral action is introduced
as a systematic approach.
The effect of the Lagrange multiplier field
and the $\pi$-flux state mean fields
is studied in Sec.\ref{sec_fluctuations},
and the dynamical mass generation is discussed.
In Sec.\ref{sec_sw}
the spin wave excitation, which is associated with phase fluctuations
of the $\pi$-flux state mean fields, is discussed.
Finally, section \ref{sec_summary} is devoted to summary and discussion.

\section{The $\pi$-Flux State and Perturbative Analysis of Fluctuations}
\label{sec_formalism}
In order to investigate the effect of fluctuations,
let us start with the Hamiltonian of the $\pi$-flux state.
Here we derive the mean field Hamiltonian and the term that describes
fluctuations about the mean field state.
The spin $1/2$ antiferromagnetic Heisenberg model on the square lattice
is given by
\begin{equation}
H=J\sum_{\langle i,j \rangle} {\bf S}_i \cdot {\bf S}_j,
\end{equation}
where $J$ is the superexchange interaction between the spins.
The summation is taken over the nearest neabor sites.
As described in Introduction, the spins are represented by
the fermions defined by eq.(\ref{eq_spin_rep})
with the constraint (\ref{eq_constraint}).
In terms of these fermions, the Hamiltonian reads,
\be
H =  - \frac{1}{2}J\sum\limits_{\left\langle {i,j} \right\rangle } 
\sum_{\alpha,\beta}
{f_{i\alpha }^\dag  f_{j\alpha } f_{j\beta }^\dag  f_{i\beta } }  
+ \sum\limits_j {\lambda _j 
\left( \sum_{\sigma} {f_{j\sigma }^\dag  f_{j\sigma }  - 1} \right)},
\ee
up to a constant term.
Here the constraint (\ref{eq_constraint}) is included by using 
a Lagrange multiplier.
Now we introduce the following mean fields,
\begin{equation}
\chi_{j,j+\delta} = \sum_{\sigma}
\langle 
f^{\dagger}_{j+\delta,\sigma} 
f_{j\sigma}
\rangle,
\end{equation}
where $\delta = \pm x, \pm y$.
The $\pi$-flux state is obtained by assuming
\begin{equation}
\chi_{j,j+{\hat x}} = \chi_{j,j-{\hat x}} = \chi,
\end{equation}
\begin{equation}
\chi_{j,j+{\hat y}} = \chi_{j,j-{\hat y}} = i \chi,
\end{equation}
and $\lambda_j = 0$.
After the Fourier transform the mean field Hamiltonian is given by
\bea
H_{MF} &=& \sum\limits_{k \in RBZ} \sum_{\alpha}
{\left( {\begin{array}{*{20}c}
   {f_{ek,\alpha }^\dag  } & {f_{ok,\alpha }^\dag  }  \\
\end{array}} \right)\left( {\begin{array}{*{20}c}
   0 & { - \kappa _k }  \\
   { - \kappa _k ^* } & 0  \\
\end{array}} \right)} \nonumber \\
& & \times
\left( {\begin{array}{*{20}c}
   {f_{ek\alpha } }  \\
   {f_{ok\alpha } }  \\
\end{array}} \right) + NJ\left| \chi  \right|^2,
\eea
where the summation with respect to $k$ is taken over
the reduced Brillouin zone, 
$|k_x - k_y|<\pi$ and $|k_x + k_y|<\pi$,
and
\be
\kappa _k  = \chi J\left( {\cos k_x  + i\cos k_y } \right).
\ee
The fields $f_{ek\sigma}$ and $f_{ok\sigma}$ are defined as
\be
f_{e\left( o \right)k\sigma }  = \sqrt {\frac{2}{N}} 
\sum\limits_{j \in A\left( B \right)} {f_{j\sigma } 
e^{ - i{\bf{k}} \cdot {\bf{R}}_j } },
\ee
where the two sublattices are labeled by $A$ and $B$.
The mean field equation is given by
\be
1 = \frac{J}{N}\sum\limits_{k \in RBZ} 
{\frac{{\cos ^2 k_x  + \cos ^2 k_y }}{{E_k }}\tanh \frac{{\beta E_k }}{2}}.
\label{eq_mfeq}
\ee
At $T=0$, we find $\chi  \simeq 0.479$.
As discussed in Introduction,
the fermion dispersion band width using this mean field value is
half of the experimentally estimated value.

In order to calculate the self-energy correction 
associated with fluctuations,
we study the Green's function.
Introducing two component spinor,
\begin{equation}
f^{\dagger}_{k\sigma} = \left(
\begin{array}{cc}
f^{\dagger}_{ek\sigma} &
f^{\dagger}_{ok\sigma} 
\end{array}
\right),
\end{equation}
we define the Matsubara Green's function as follows,
\be
G_{k\sigma} (\tau) = -\langle f_{k\sigma} (\tau)
f_{k\sigma}^{\dagger}(0) \rangle.
\ee
In the mean field state the Green's function is
\be
G^{(0)}_{k \uparrow } \left( {i\omega _n } \right) = \left( {\begin{array}{*{20}c}
   {\frac{{u_k^2 }}{{i\omega _n  + E_k }} + \frac{{\left| {v_k } \right|^2 }}{{i\omega _n  - E_k }}} & {u_k v_k \left( {\frac{1}{{i\omega _n  + E_k }} - \frac{1}{{i\omega _n  - E_k }}} \right)}  \\
   {u_k v_k^* \left( {\frac{1}{{i\omega _n  + E_k }} - \frac{1}{{i\omega _n  - E_k }}} \right)} & {\frac{{\left| {v_k } \right|^2 }}{{i\omega _n  + E_k }} + \frac{{u_k^2 }}{{i\omega _n  - E_k }}}  \\
\end{array}} \right),
\ee
\be
G^{(0)}_{k \downarrow } \left( {i\omega _n } \right) = \left( {\begin{array}{*{20}c}
   {\frac{{u_k^2 }}{{i\omega _n  + E_k }} + \frac{{\left| {v_k } \right|^2 }}{{i\omega _n  - E_k }}} & { - u_k v_k \left( {\frac{1}{{i\omega _n  + E_k }} - \frac{1}{{i\omega _n  - E_k }}} \right)}  \\
   { - u_k v_k^* \left( {\frac{1}{{i\omega _n  + E_k }} - \frac{1}{{i\omega _n  - E_k }}} \right)} & {\frac{{\left| {v_k } \right|^2 }}{{i\omega _n  + E_k }} + \frac{{u_k^2 }}{{i\omega _n  - E_k }}}  \\
\end{array}} \right),
\ee
where 
$u_k  = 1/\sqrt{2}$
and $v_k = {\kappa _k}/(\sqrt{2} {\left| {\kappa _k } \right|})$.

\begin{figure}
   \begin{center}
    \includegraphics[width=2in]{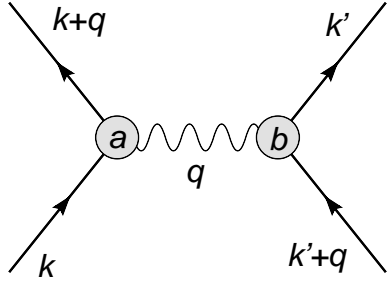}
   \end{center}
   \caption{ \label{fig_vertices}
	The Feynman diagram for the interaction term associated with 
	fluctuations of $\chi_{j,j+\delta}$.
	Here $(a,b) = 
	(\tau_o \sigma_{\uparrow}, \tau_e \sigma_{\uparrow}),
	(\tau_o \sigma_{\downarrow}, \tau_e \sigma_{\downarrow}),
	(\tau_o \sigma_-, \tau_e \sigma_+),
	(\tau_o \sigma_+, \tau_e \sigma_-)$.
    }
 \end{figure}

Having defined the mean field Green's function,
let us derive the term describing fluctuations.
Fluctuations about the mean field, $\chi$,
arise from
\bea
H_{{\mathop{\rm int}} }  &=& \frac{{4J}}{N}\sum\limits_{k,k',q \in RBZ, q\neq 0} 
\left( {\cos q_x  + \cos q_y } \right)
\nonumber \\
& & \times
f_{o,k + q,\alpha }^\dag  
f_{ok\beta } f_{ek'\beta }^\dag  f_{ek' + q\alpha } . 
\eea
Note that $q=0$ contribution is excluded.
Because the $q=0$ term is included in the mean field Hamiltonian.
In order to make the notation simpler, we omit $q\neq 0$ in
the following equations.
The interaction vertices are represented in Fig. \ref{fig_vertices}.
For example, the diagram with $a=\tau_o \sigma_+$ and $b=\tau_e \sigma_-$ 
in Fig.\ref{fig_vertices} corresponds to the following term,
\bea
H_{(\tau_o \sigma_+,\tau_e \sigma_-)}  
&=& \frac{{4J}}{N}\sum\limits_{k,k',q \in RBZ} 
\left( {\cos q_x  + \cos q_y } \right) \nonumber \\
& & \times f_{o,k + q,\uparrow }^\dag  
f_{ok\downarrow } f_{ek'\downarrow }^\dag  f_{ek' + q\uparrow }. 
\eea

The importance of including the fluctuation effect
is clearly demonstrated by the lowest order self-energy correction.
The first order self-energy with respect to $H_{\rm int}$ 
for the spin-up fermions is
\begin{equation}
\begin{array}{c}
 \Sigma _{k \uparrow } \left( {i\omega _n } \right) 
=  - \frac{1}{{\beta N}}\sum\limits_{q \in RBZ} {\sum\limits_{i\Omega _n } 
{\sum\limits_{\sigma} {2J_q } } \left( {\begin{array}{*{20}c}
   0 & {}  \\
   {} & 1  \\
\end{array}} \right)} G_{k + q\sigma } \left( {i\omega _n  + i\Omega _n } \right)\left( {\begin{array}{*{20}c}
   1 & {}  \\
   {} & 0  \\
\end{array}} \right) \\ 
  - \frac{1}{{\beta N}}\sum\limits_{q \in RBZ} {\sum\limits_{i\Omega _n } 
{\sum\limits_{\sigma} {2J_q } } \left( {\begin{array}{*{20}c}
   1 & {}  \\
   {} & 0  \\
\end{array}} \right)} G_{k + q\sigma } \left( {i\omega _n  + i\Omega _n } \right)\left( {\begin{array}{*{20}c}
   0 & {}  \\
   {} & 1  \\
\end{array}} \right) \\ 
 \end{array}.
\end{equation}
Straightforward calculation leads to 
\bea
\Sigma _{k \uparrow } \left( {i\omega _n } \right) 
&=&  - \frac{2}{N}\sum\limits_{k' \in RBZ} {J_{k' - k} } \left( {\begin{array}{*{20}c}
   0 & {2u_{k'} v_{k'} }  \\
   {2u_{k'} v_{k'}^* } & 0  \\
\end{array}} \right) \nonumber \\
&=& \left( {\begin{array}{*{20}c}
   0 & { - CJ\cos k_x  - iCJ\cos k_y }  \\
   { - CJ\cos k_x  + iCJ\cos k_y } & 0  \\
\end{array}} \right),
\eea
with $C\simeq 0.479$.
Thus, the band width of the quasiparticle dispersion,
which is given by $E_k^{MF}  = \chi J\sqrt {\cos ^2 k_x  + \cos ^2 k_y }$,
is exactly doubled by this self-energy.
Therefore, the quantitative difference of the quasiparticle band width compared 
to the experiment is much improved by including the fluctuation effect.

\section{Coherent State Path-Integral Formalism}
\label{sec_cpi}
In principle, one can extend perturbative analysis 
in the previous section
to higher order trerms.
However, since there are eight vertices, drawing Feynman diagrams
is complicated.
Furthermore, the Hamiltonian formalism is not useful 
for the study of fluctuations of the Lagrange multiplier field.
For this purpose, the coherent state path-integral formulation
is convenient.
In the coherent state path-integral formulation, the partition function
is given by
\be
\Xi  = \int {Df^\dag  DfD\lambda } \exp \left[ { - S} \right],
\ee
where the action is
\be
S = \int_0^\beta  {d\tau } \left[ {\sum\limits_j {f_{j\sigma }^\dag  
\left( {\partial _\tau   + \lambda _j } \right)f_{j\sigma } }  
- \frac{1}{2}J\sum\limits_{\left\langle {i,j} \right\rangle } 
{f_{i\alpha }^\dag  f_{j\alpha } f_{j\beta }^\dag  f_{i\beta } } } \right],
\ee
with $\beta=1/k_B T$ the inverse temperature.
After the Fourier transform, the Lagrangian is given by
\be
\begin{array}{c}
 L = \sum\limits_{k \in RBZ} 
  {
	\left(
	f_{ek\sigma }^\dag  \partial _\tau  f_{ek\sigma } 
       + f_{ok\sigma }^\dag  \partial _\tau  f_{ok\sigma }
	\right)
  }  
- \frac{J}{N}\sum\limits_{k,k',q \in RBZ} 
{\sum\limits_{\delta  =  \pm x , \pm y } {f_{ok + q\alpha }^\dag  
f_{ek\alpha } f_{ek'\beta }^\dag  f_{ok' + q\beta } } } 
e^{ - i\left( {k - k'} \right) \cdot \delta }  \\ 
  + \sum\limits_{k,q \in RBZ} 
{\left( {\lambda _q f_{ek + q\sigma }^\dag  f_{ek\sigma }  
+ \lambda _{q + Q} f_{ek + q\sigma }^\dag  f_{ek\sigma }  
+ \lambda _q f_{ok + q\sigma }^\dag  f_{ok\sigma }  
- \lambda _{q + Q} f_{ok + q\sigma }^\dag  f_{ok\sigma } } \right)}  \\ 
 \end{array},
\ee
where $Q=(\pi,\pi)$.

Now we introduce a Storatonovich-Hubbard transformation to rewrite the
interaction term as follows,
\begin{eqnarray*}
L &=& \sum\limits_{k \in RBZ} {\left( {f_{ek\sigma }^\dag  \partial _\tau  f_{ek\sigma }  
+ f_{ok\sigma }^\dag  \partial _\tau  f_{ok\sigma } } \right)}  \\ 
& &
+ \sum\limits_{q \in RBZ} {\sum\limits_{k \in RBZ} 
{\left( {\lambda _q f_{ek + q\sigma }^\dag  f_{ek\sigma }  
+ \lambda _{q + Q} f_{ek + q\sigma }^\dag  f_{ek\sigma }  
+ \lambda _q f_{ok + q\sigma }^\dag  f_{ok\sigma }  
- \lambda _{q + Q} f_{ok + q\sigma }^\dag  f_{ok\sigma } } \right)} }  \\ 
& &
  - \frac{J}{2}\sum\limits_{q \in RBZ} {\sum\limits_{\delta  =  \pm x , \pm y } 
{\chi _q^{\left( \delta  \right)*} \left( {\sqrt {\frac{2}{N}} 
\sum\limits_{k \in RBZ} {e^{ik \cdot \delta } f_{ek\alpha }^\dag  
f_{ok + q\alpha } } } \right)} }  \\ 
& &
  - \frac{J}{2}\sum\limits_{q \in RBZ} {\sum\limits_{\delta  =  \pm x , \pm y } 
{\chi _q^{\left( \delta  \right)} \left( {\sqrt {\frac{2}{N}} \sum\limits_{k \in RBZ} 
{e^{ - ik \cdot \delta } f_{ok + q\alpha }^\dag  f_{ek\alpha } } } \right)} }  \\ 
& &
  + \frac{J}{2}\sum\limits_{q \in RBZ} {\sum\limits_{\delta  =  \pm x , \pm y } 
{\chi _q^{\left( \delta  \right)*} \left( {\chi _q^{\left( \delta  \right)} } \right)} }
\end{eqnarray*}
At the saddle point, we see that
\be
\chi _q^{\left( \delta  \right)}  
= \sqrt {\frac{2}{N}} \sum\limits_{k \in RBZ} 
{e^{ik \cdot \delta } \left\langle {f_{ek\alpha }^\dag  
f_{ok + q\alpha } } \right\rangle }.
\ee
As a mean field state, we assume that
\begin{eqnarray*}
\chi _q^{\left( { + x } \right)}  &=& \sqrt {\frac{N}{2}} \chi _1 \delta _{q,0} , 
\hspace{1em}
\chi _q^{\left( { + y } \right)} = \sqrt {\frac{N}{2}} \chi _2 \delta _{q,0} , 
\\
\chi _q^{\left( { - x } \right)} &=& \sqrt {\frac{N}{2}} \chi _3 \delta _{q,0} , 
\hspace{1em}
\chi _q^{\left( { - y } \right)} = \sqrt {\frac{N}{2}} \chi _4 \delta _{q,0},
\end{eqnarray*}
\be
\lambda _q  = 0, \hspace{1em} \lambda _{q + Q}  = 0.
\ee
The action is
\bea
S &=& \int_0^\beta  {d\tau } 
\left[ \sum\limits_{k \in RBZ} 
 \left( {\begin{array}{*{20}c}
   {f_{ek\alpha }^\dag  } & {f_{ok\alpha }^\dag  }  \\
\end{array}} \right)\left( {\begin{array}{*{20}c}
   {\partial _\tau  } & {\kappa _k^* }  \\
   {\kappa _k } & {\partial _\tau  }  \\
\end{array}} \right)
   \left( {\begin{array}{*{20}c}
     {f_{ek\alpha } }  \\
     {f_{ok\alpha } }  \\
   \end{array}} \right) \right.
\nonumber \\
& & \left. 
+ \frac{{NJ}}{4}\left( 
{\left| {\chi _1 } \right|^2  + \left| {\chi _2 } \right|^2  
+ \left| {\chi _3 } \right|^2  + \left| {\chi _4 } \right|^2 } \right) 
\right],
\eea
where
\be
\kappa _k  =  - \frac{J}{2}\left( {\chi _1 e^{ - ik_x }  
+ \chi _2 e^{ - ik_y }  + \chi _3 e^{ik_x }  + \chi _4 e^{ik_y } } \right).
\ee
For the $\pi$-flux state, $\chi_j$'s are
\be
\chi_1=\chi_3=-i\chi_2=-i\chi_4\equiv \chi.
\ee
The saddle point equation, or the mean field equation, is given by
eq. (\ref{eq_mfeq}).

\section{Effect of Fluctuations}
\label{sec_fluctuations}
Now we study the fluctuations of the Lagrange multiplier field
and the $\pi$-flux state mean fields, $\chi$.
Fluctuations about the saddle point are described by the following action,
\bea
S_f  &=& \int_0^\beta  {d\tau } 
\left[ {\frac{J}{2}\sum\limits_{q \in RBZ} {\sum\limits_{\delta  =  \pm x , \pm y } 
{\left| {\chi _{q\delta } } \right|^2 } } } 
+ \sum\limits_{k,q \in RBZ} \left( {\begin{array}{*{20}c}
   {f_{ek + q\sigma }^\dag  } & {f_{ok + q\sigma }^\dag  } \\
 \end{array}} \right) 
\right. \nonumber \\ 
& & \left.  
\times \left( {\begin{array}{*{20}c}
   {\lambda _{eq} } & { - \frac{J}{{\sqrt {2N} }}
\sum\limits_{\delta  =  \pm x , \pm y } {e^{i\left( {k + \frac{q}{2}} \right)
 \cdot \delta } \chi _{ - q,\delta }^* } }  \\
   { - \frac{J}{{\sqrt {2N} }}\sum\limits_{\delta  =  \pm x , \pm y } 
{e^{ - i\left( {k + \frac{q}{2}} \right) \cdot \delta } \chi _{q,\delta } } }
& {\lambda _{oq} }  \\
\end{array}} \right)\left( {\begin{array}{*{20}c}
   {f_{ek\sigma } }  \\
   {f_{ok\sigma } }  \\
\end{array}} \right)  \right],
\eea
where 
$\chi _{q,\delta }  = e^{iq \cdot \delta /2} \chi _q^{\left( \delta  \right)}$
and $\lambda_{eq}=\lambda_q+\lambda_{q+Q}$
and $\lambda_{oq}=\lambda_q-\lambda_{q+Q}$.
We consider fluctuations within the Gaussian approximation.
At this point, we include the staggered magnetization term:
\be
S_{st}  = \int_0^\beta  {d\tau } \sum\limits_{k \in RBZ} {\left( {\begin{array}{*{20}c}
   {f_{ek}^\dag  } & {f_{ok}^\dag  }  \\
\end{array}} \right)\left( {\begin{array}{*{20}c}
   { - \Delta _{st} \sigma _z } & 0  \\
   0 & {\Delta _{st} \sigma _z }  \\
\end{array}} \right)\left( {\begin{array}{*{20}c}
   {f_{ek} }  \\
   {f_{ok} }  \\
\end{array}} \right)}.
\ee
Within the mean field theory, $\Delta_{st}=0$.
Nonzero value of $\Delta_{st}$ arises from the self-energy effect 
as shall be seen below.
Integrating out the fermion fields leads to the following effective action,
\bea
S_f  &=&  - \frac{1}{8}\sum\limits_{q} {\left( {\begin{array}{*{20}c}
   {\lambda _{eq} } & {\lambda _{oq} }  \\
\end{array}} \right)\left( {\begin{array}{*{20}c}
   {\Pi _0^\lambda  \left( q \right)} & { - \Pi _1^\lambda  \left( q \right)}  \\
   { - \Pi _1^\lambda  \left( q \right)} & {\Pi _0^\lambda  \left( q \right)}  \\
\end{array}} \right)\left( {\begin{array}{*{20}c}
   {\lambda _{e, - q} }  \\
   {\lambda _{o, - q} }  \\
\end{array}} \right)}  \nonumber \\ 
& &  + \frac{J}{2}\sum\limits_q {\sum\limits_{\delta  =  \pm x , \pm y } 
	{\left( {\begin{array}{*{20}c}
   {\chi _{q\delta }^* } & {\chi _{ - q,\delta } }  \\
\end{array}} \right)\left( {\begin{array}{*{20}c}
   {1 - \Pi _0^\chi  \left( q \right)} & {\Pi _1^\chi  \left( q \right)}  \\
   {\Pi _2^\chi  \left( q \right)} & {1 - \Pi _0^\chi  \left( q \right)}  \\
\end{array}} \right)\left( {\begin{array}{*{20}c}
   {\chi _{q\delta } }  \\
   {\chi _{ - q,\delta }^* }  \\
\end{array}} \right)} },
\eea
where
\be
\Pi _0^\lambda  \left( q \right) = \sum\limits_{{\bf k}\in RBZ} 
{\left( {\frac{1}{{i\Omega _n  + E_{{\bf k} + {\bf q}}  + E_{\bf k} }} 
- \frac{1}{{i\Omega _n  - E_{{\bf k} + {\bf q}}  - E_{\bf k} }}} \right)},
\ee
\be
\Pi _1^\lambda  \left( q \right) = \sum\limits_{{\bf k} \in RBZ}  
{\frac{{\kappa _{\bf k}^* \kappa _{{\bf k} + {\bf q}}  + \Delta _{st}^2 }}{{E_{\bf k} E_{{\bf k} + {\bf q}} }}
\left( {\frac{1}{{i\Omega _n  + E_{{\bf k} + {\bf q}}  + E_{\bf k} }} - \frac{1}{{i\Omega _n  - E_{{\bf k} + {\bf q}}  - E_{\bf k} }}} \right)},
\ee
\be
\Pi _0^\chi  \left( q \right) = \frac{J}{{8N}}\sum\limits_{{\bf k} \in RBZ} {\left( {\frac{1}{{i\Omega _n  + E_{{\bf k} + {\bf q}}  + E_{\bf k} }} - \frac{1}{{i\Omega _n  - E_{{\bf k} + {\bf q}}  - E_{\bf k} }}} \right)},
\ee
\be
\Pi _1^\chi  \left( q \right) = \frac{J}{N}\sum\limits_{{\bf k} \in RBZ}  
{\frac{{\kappa _{\bf k} \kappa _{{\bf k} + {\bf q}}  - \Delta _{st}^2 }}{{E_{\bf k} E_{{\bf k} + {\bf q}} }}\left( {\frac{1}{{i\Omega _n  + E_{{\bf k} + {\bf q}}  + E_{\bf k} }} - \frac{1}{{i\Omega _n  - E_{{\bf k} + {\bf q}}  - E_{\bf k} }}} \right)},
\ee
\be
\Pi _2^\chi  \left( q \right) = \frac{J}{N}\sum\limits_{{\bf k} \in RBZ}  
{\frac{{\kappa _{\bf k}^* \kappa _{{\bf k} + {\bf q}}^*  - \Delta _{st}^2 }}{{E_{\bf k} E_{{\bf k} + {\bf q}} }}\left( {\frac{1}{{i\Omega _n  + E_{{\bf k} + {\bf q}}  + E_{\bf k} }} - \frac{1}{{i\Omega _n  - E_{{\bf k} + {\bf q}}  - E_{\bf k} }}} \right)}.
\ee
Here $E_{\bf k}=\sqrt{
\chi^2 J^2 \left( \cos^2 k_x + \cos^2 k_y \right) + \Delta_{st}^2}$,
and 
$q$ denotes $({\bf q},i\Omega_n)$ with ${\bf q} \in $RBZ and $\Omega_n = 2\pi n/\beta$
being the bosonic Matsubara frequency.

\begin{figure}
   \begin{center}
    \includegraphics[width=3.4in]{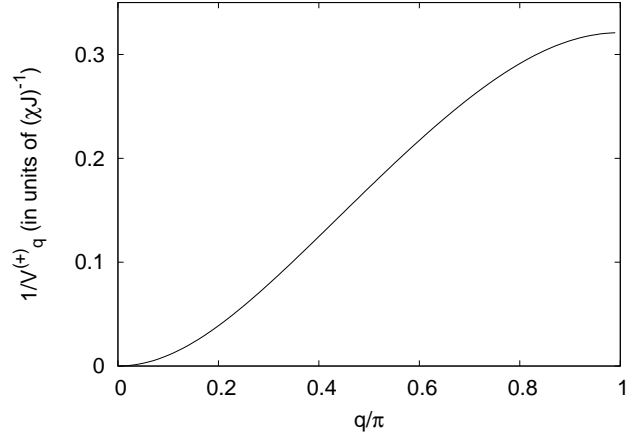}
   \end{center}
   \caption{ \label{fig_Vqp}
	The inverse of the potential $V_{{\bf q}=(q,q)}^{(+)}$ versus $q/\pi$. 
    }
 \end{figure}

In order to make clear the physical meaning of $\lambda_q$ and $\chi_{q\delta}$,
we study them separately. 
First, let us investigate the effect of the former.
As we will see below, exchange of $\lambda_q$ fields leads to 
a logarithmic confining potential between the fermions
in the presence of the fermion mass term.
The coupling term between $\lambda_q$ and the fermions is
\bea
S_\lambda ^{{\mathop{\rm int}} }  
&=& \sum\limits_{k,q} {\left[ {\frac{1}{2}\left( {\begin{array}{*{20}c}
   {\lambda _{eq} } & {\lambda _{oq} }  \\
\end{array}} \right)\left( {\begin{array}{*{20}c}
   {f_{e{\bf k}+{\bf q}\sigma }^\dag  f_{e{\bf k}\sigma } }  \\
   {f_{o{\bf k} + q\sigma }^\dag  f_{o{\bf k}\sigma } }  \\
\end{array}} \right) + \frac{1}{2}\left( {\begin{array}{*{20}c}
   {f_{e{\bf k}\sigma }^\dag  f_{e{\bf k}+{\bf q}\sigma } } & 
{f_{o{\bf k}\sigma }^\dag  f_{o{\bf k} + q\sigma } }  \\
\end{array}} \right)\left( {\begin{array}{*{20}c}
   {\lambda _{e, - q} }  \\
   {\lambda _{o, - q} }  \\
\end{array}} \right)} \right]}  
\nonumber \\
& & 
  - \frac{1}{8}\sum\limits_q {\left( {\begin{array}{*{20}c}
   {\lambda _{eq} } & {\lambda _{oq} }  \\
\end{array}} \right)\left( {\begin{array}{*{20}c}
   {\Pi _0^\lambda  \left( q \right)} & { - \Pi _1^\lambda  \left( q \right)}  \\
   { - \Pi _1^\lambda  \left( q \right)} & {\Pi _0^\lambda  \left( q \right)}  \\
\end{array}} \right)\left( {\begin{array}{*{20}c}
   {\lambda _{e, - q} }  \\
   {\lambda _{o, - q} }  \\
\end{array}} \right)}.
\label{eq_Seff}
\eea
Note that using this action in deriving the effective interaction term
is equivalent to the random phase approximation.
This point is demonstrated using a simple model in Appendix.
Integrating out $\lambda_q$ leads to the following interaction term,
\bea
S_\lambda   &=& \sum\limits_q {\left( {\begin{array}{*{20}c}
   {\rho _{eq}  + \rho _{oq} } & {\rho _{eq}  - \rho _{oq} }  \\
\end{array}} \right)} \left( {\begin{array}{*{20}c}
   {V_q^{\left(  +  \right)} } & 0  \\
   0 & {V_q^{\left(  -  \right)} }  \\
\end{array}} \right) \nonumber \\
& & \times \left( {\begin{array}{*{20}c}
   {\rho _{e, - q}  + \rho _{o, - q} }  \\
   {\rho _{e, - q}  - \rho _{o, - q} }  \\
\end{array}} \right),
\eea
where
\be
V_q^{\left(  \pm  \right)}  = 
\frac{1}{{\Pi _{0}^\lambda  \left( q \right) \mp \Pi _{1}^\lambda  \left( q \right)}}.
\ee
One can see that for the static case, $i\Omega_n=0$,
the denominator of $V_q^{(+)}$ behaves as 
$1/q^2$ for $q \rightarrow 0$ as seen from the expansion
with respect to $q$.
($1/V_q^{(+)}$ is shown in Fig.\ref{fig_Vqp}.)
Therefore, the potential term $V_q^{(+)}$ is a logarithmic potential.
\cite{Laughlin1995}
Since such a potential diverges at long distance,
it is a kind of confining potential.
The same instantaneous interaction term is obtained in the continuum theory of QED$_3$
by taking the Coulomb gauge.

Now we examine dynamical mass generation associated with 
this logarithmic confining potential.
We consider the first order self-energy term.
The equation to determine the self-energy in a self-consistent manner is
\be
\Sigma _{e{\bf k}} \left( {i\omega _n } \right) 
=  - \frac{1}{{\beta N}}\sum\limits_q {V_q^{\left(  +  \right)} 
G_{e{\bf k}+{\bf q}} \left( {i\omega _n  + i\Omega _n } \right)}.
\label{eq_sfe}
\ee
A similar equation holds for $\Sigma_{o{\bf k}}  \left( {i\omega _n } \right)$.
Here $G_{e{\bf k}}(i\omega_n ) 
= G^{(0)}_{e{\bf k}}(i\omega_n) - \Sigma_{e{\bf k}} (i\omega_n )$.
To study the dynamical mass generation, 
we need to consider a non-uniform mass term in general.
However, for such a general mass term we are unable to obtain analytic forms 
for the propagators.
To solve the equation (\ref{eq_sfe})
self-consistently we need to solve the coupled equations
numerically.
Here we discuss the dynamical mass generation within the approximation of 
the uniform mass term, 
$\Sigma_{e{\bf k}} (i\omega_n ) = -\Delta_{st} \sigma_z$.
Under this approximation, we find the equation for the mass,
\be
\frac{1}{N}\sum\limits_{{\bf q}}
{\frac{{V_{\bf q}^{\left(  +  \right)} }}{{2E_{(\pi/2,\pi/2) + {\bf q}} }} = 1}.
\ee
In spite of this simple form,
numerically solving this gap equation is not an easy task.
Because the potential is singular, it is necessary to introduce 
an infrared cutoff and the result depends on it.
Furthermore, the result also depends on the number of lattice 
points introduced for numerical estimations.\cite{Gusynin2003}
Instead of precisely determining the mass gap value,
we evaluate it approximately.
After some algebra, we find
\be
\Pi _{ee}^\lambda  \left( q \right) - \Pi _{eo}^\lambda  \left( q \right) 
= \frac{1}{N}\sum\limits_{{\bf k} \in RBZ} 
{\frac{{A_{{\bf k},{\bf q}}^2 }}{{E_{\bf k}^3 }} + } 
\left( {\sin ^2 \frac{{q_x }}{2} + \sin ^2 \frac{{q_y }}{2}} \right)
\frac{1}{N}\sum\limits_{{\bf k} \in RBZ} 
{\frac{{\left( {\chi J} \right)^2 }}{{E_{\bf k}^3 }}},
\label{eq_pp}
\ee
where
\be
A_{{\bf k},{\bf q}}^2  \simeq \frac{1}{4}q_\alpha  q_\beta  
\left( {\frac{{\partial ^2 E_{\bf k} }}{{\partial k_\alpha  \partial k_\beta  }}
E_{\bf k}  - \frac{{\partial E_{\bf k} }}{{\partial k_\alpha  }}\frac{{\partial E_{\bf k} }}{{\partial k_\beta  }}} \right).
\ee
In the $k$-summation in eq.(\ref{eq_pp}), 
dominant contribution comes from ${\bf k} \simeq (\pm \pi/2, \pm \pi/2)$.
Taking the approximate form around these points, we obtain
\be
\Pi _{ee}^\lambda  \left( q \right) - \Pi _{eo}^\lambda  \left( q \right) 
\simeq \frac{1}{2}\frac{{\left( {\chi J} \right)^2 }}{{\Delta _{st}^3 }}{\bf q}^2.
\ee
Now the gap equation is
\be
\int_0^{\pi} \frac{d q}{2\pi}
\frac{\Delta_{st}^2}{(\chi J)^2 q} \simeq 1.
\label{eq_gap}
\ee
Note that this equation is different from that for QED$_3$
by the factor of $(\Delta_{st}/\chi J)^2$ in which
the potential term is not proportional to $\Delta_{st}$.
Apparently this gap equation is suffering from the infrared divergence.
In order to properly deal with this infrared divergence,
we need to take into account the vertex correction.
In the QED$_3$ theory
there is controversy in the choice of the vertex correction.
\cite{Kondo1992,Aitchison1996}
Here we determine it from a physical argument.
The interaction term in the gap equation comes from the following term,
\be
H_{\rm int} = \sum_{q} V_q^{(+)} \delta \rho_q \delta \rho_{-q},
\ee
where $\delta \rho_q = \rho_q - \overline{\rho}$,
with $\overline{\rho}$ being the 
uniformely distributed background particle density.
Note that $\delta \rho_q \rightarrow 0$ for $q\rightarrow 0$.
On the other hand, from the density-density correlation function,
$\langle \delta \rho_q \delta \rho_{-q} \rangle \sim q^2$.
This suggests that the approximate form of the vertex correction is 
\be
\gamma _q  = \left\{ \begin{array}{l}
 \left( {\Delta _{st} /\chi J} \right)^{ - \alpha /2} q^\alpha   
  \hspace{2em} \left( {\rm for} \hspace{1em} {q < \sqrt {\Delta _{st} /\chi J} } \right) \\ 
  1  \hspace{2em} 
\left( {\rm for} \hspace{1em} {q > \sqrt {\Delta _{st} /\chi J} } \right) \\ 
 \end{array} \right.,
\ee
with $\alpha=1$.	
Dividing the integration with respect to $q$ in eq.(\ref{eq_gap})
with including $\gamma_q$, into $0<q<\sqrt(\Delta_{st}/\chi J)$ and
$q>\sqrt(\Delta_{st}/\chi J)$, 
\be
2\pi (\Delta_{st}/\chi J)^2  - 
\log \left( {\frac{\pi }{(\Delta_{st}/\chi J)}} \right) \simeq 1.
\ee
This equation can be solved numerically, and the solution is
\be
\Delta_{st} \simeq 0.616 \chi J.
\ee
Substituting $\chi \simeq 2 \times 0.479$, we find $\Delta_{st} \simeq 0.59J$.
This value corresponds to the staggered magnetization of $m_{st} \simeq 0.30$,
which is a reasonable value compared to the numerical simulation results
and the experiments.

Having discussed the dynamical mass generation associated with
the Lagrange multiplier field fluctuations,
let us move on to the mean field fluctuations, $\chi_{q\delta}$.
The fluctuation mode associated with $\chi_{q\delta}$ is
found from the pole of the following equation,
\be
\det \left( {\begin{array}{*{20}c}
   {1 - \Pi _0^\chi  \left( q \right)} & {\Pi _1^\chi  \left( q \right)}  \\
   {\Pi _2^\chi  \left( q \right)} & {1 - \Pi _0^\chi  \left( q \right)}  \\
\end{array}} \right) = 0.
\ee
The pole was found from numerical computations 
with mesh size of $64 \times 64$ and taking $\delta=0.10$,
where $\delta$ is introdued in the analytic continuation 
from the Matsubara frequency, $i\Omega_n$ to
the real frequency, through $i\Omega_n = \omega + i\delta$.
It was found that the amplitude fluctuation mode has
the excitation energy gaps whose minimum is about $1.2J$.
Along the line from $(0,0)$ to $(\pi/2,\pi/2)$ in the 
reduced Brillouin zone,
the band minimum is located at $(0,0)$,
and the maximum is at $(\pi/2,\pi/2)$.
The band width along this line is $1.3J$.
The excitation energy at other points are within this energy range.
Although the amplitude fluctuation mode of $\chi_{q\delta}$ is
high-energy mode and can be neglected,
the phase fluctuation of $\chi_{q\delta}$ is related to
the spin-wave excitation.
Because there is the relation, 
\be
\left\langle {{\bf{S}}_j  \cdot {\bf{S}}_{j + \delta } } \right\rangle  
= \left| {\chi _{j,j + \delta } } \right|^2  
= \left| {\sqrt {\frac{2}{N}} 
\sum\limits_{{\bf{q}} \in RBZ} {e^{i{\bf{q}} 
\cdot \left( {{\bf{R}}_j  + \delta /2} \right)} } \chi _{q,\delta } } \right|^2.
\ee
The spin wave excitation is discussed in the next section.

\section{Spin-Wave Excitations}
\label{sec_sw}
In this section, we consider the spin wave excitation 
associated with the phase fluctuations of the mean field, $\chi_{q\delta}$
in the presence of the dynamically generated mass, or the staggered magnetization.
The calculation is simiar to the spin-density wave theory.
\cite{Schrieffer1989,Singh1990}
Here we calculate the spin wave excitation spectrum 
following ref.\citenum{Chi1992}.

\begin{figure}
   \begin{center}
    \includegraphics[width=3.4in]{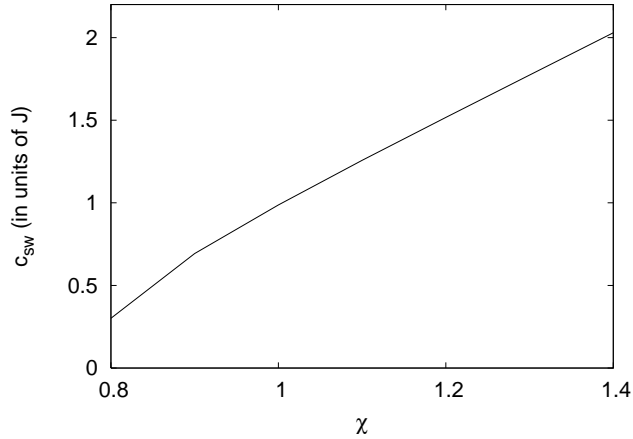}
   \end{center}
   \caption{ \label{fig_chi_vs_csw}
	The mean field $\chi$ dependence of the spin-wave velocity, $c_{sw}$.
    }
 \end{figure}

The repulsive interaction $V_q^{(+)}$ 
that leads to the non-zero staggered magnetization
is approximated by a short-range repulsion, $V$, for simplicity.
The interaction $V$ is associated with a weak short-range repulsive interaction 
assumed by Hsu.\cite{Hsu1990}
The value of $V$ is evaluated from 
\[
\frac{1}{N}\sum\limits_{k \in RBZ} 
{\frac{1}{{\sqrt {\left| {\kappa _{\bf k} } \right|^2  
+ \left| {\Delta _{st} } \right|^2 } }}}  
= \frac{1}{V}.
\]
Using the values of $\chi=0.994$ and $\Delta_{st}=0.616J$,
we obtain $V=2.2J$. 
The action that describes fluctuations is given by
\bea
S &=& \int_0^\beta  {d\tau } \sum\limits_{k} {\left( {\begin{array}{*{20}c}
   {f_k^\dag  } & {f_{k + Q}^\dag  }  \\
\end{array}} \right)\left( {\begin{array}{*{20}c}
   {\partial _\tau   + \chi J\cos k_x } & { - i\chi J\cos k_y  
- \Delta _{st} \sigma _z }  \\
   {i\chi J\cos k_y  - \Delta _{st} \sigma _z } & {\partial _\tau   - \chi J\cos k_x }  
\\
\end{array}} \right)\left( {\begin{array}{*{20}c}
   {f_k}  \\
   {f_{k + Q}}  \\
\end{array}} \right)} \nonumber \\
& & 
+ \int_0^\beta  {d\tau } \left[ { - \sum\limits_{k,k'} 
{\delta {\bf{m}}_{k - k'} \cdot \left( {f_k}^\dag  \sigma f_{k'}  \right)}
+ \frac{1}{V}\sum\limits_q {\delta {\bf{m}}_q  \cdot \delta {\bf{m}}_{ - q} } } \right] 
+ \frac{{\beta N}}{V}\Delta _{st}^2.
\eea
Integrating out fermions, and after some algebra, we obtain
\bea
S_{eff} &=& \frac{1}{V}\sum\limits_{q \in RBZ} 
{\left( {a_q \delta m_q^z \delta m_{ - q}^z  + a_{q + Q} 
\delta m_{q + Q}^z \delta m_{ - q - Q}^z } \right)} 
\nonumber \\ 
& & + \frac{1}{V}\sum\limits_{q \in RBZ} {\left( {\begin{array}{*{20}c}
   {\delta m_q^x } & {\delta m_{q + Q}^y }  \\
\end{array}} \right)\left( {\begin{array}{*{20}c}
   {b_q } & {c_q }  \\
   { - c_q } & {b_{q + Q} }  \\
\end{array}} \right)\left( {\begin{array}{*{20}c}
   {\delta m_{ - q}^x }  \\
   {\delta m_{ - q - Q}^y }  \\
\end{array}} \right)} \nonumber \\
& &  + \frac{1}{V}\sum\limits_{q \in RBZ} {\left( {\begin{array}{*{20}c}
   {\delta m_{q + Q}^x } & {\delta m_q^y }  \\
\end{array}} \right)\left( {\begin{array}{*{20}c}
   {b_q } & { - c_q }  \\
   {c_q } & {b_{q + Q} }  \\
\end{array}} \right)\left( {\begin{array}{*{20}c}
   {\delta m_{ - q - Q}^x }  \\
   {\delta m_{ - q}^y }  \\
\end{array}} \right)},
\eea
where 
\be
a_q  = 1 - \frac{1}{2}V\left[ {K_q^{\left( 0 \right)}  
- K_q^{\left( z \right)} } \right], b_q  = 
1 - \frac{1}{2}VK_q^{\left( 0 \right)} , 
c_q  = \frac{1}{2}VK_q^{\left( u \right)},
\ee
with 
\be
K_q^{\left( 0 \right)}  = \frac{1}{N}
 \sum\limits_{{\bf k} \in RBZ} 
{\left( {1 - \frac{{\kappa _{k + q}^* \kappa _{\bf k}  
+ \kappa _{k + q} \kappa _{\bf k}^*  - 2\Delta _{st}^2 }}{{2E_{{\bf k} + {\bf q}} E_{\bf k} }}} \right)} 
\left( {\frac{1}{{i\Omega _n  + E_{{\bf k} + {\bf q}}  + E_{\bf k} }} 
- \frac{1}{{i\Omega _n  - E_{{\bf k} + {\bf q}}  - E_{\bf k} }}} \right),
\ee
\be
K_q^{\left( z \right)}  = \frac{1}{N} \sum\limits_{{\bf k} \in RBZ} 
{\frac{{2\Delta _{st}^2 }}{{E_{{\bf k} + {\bf q}} E_{\bf k} }}} 
\left( {\frac{1}{{i\Omega _n  + E_{{\bf k} + {\bf q}}  + E_{\bf k} }} 
- \frac{1}{{i\Omega _n  - E_{{\bf k} + {\bf q}}  - E_{\bf k} }}} \right),
\ee
\be
K_q^{\left( u \right)}  = \frac{1}{N} \sum\limits_{{\bf k} \in RBZ} 
{\left( { - i\Delta _{st} } \right)} \left( {\frac{1}{{E_{\bf k} }} 
+ \frac{1}{{E_{{\bf k} + {\bf q}} }}} \right)\left( {\frac{1}{{i\Omega _n  
+ E_{{\bf k} + {\bf q}}  + E_{\bf k} }} + \frac{1}{{i\Omega _n  
- E_{{\bf k} + {\bf q}}  - E_{\bf k} }}} \right).
\ee
The spin wave excitaion is associated with the pole of 
the transverse spin fluctuations.
The pole is found from
\be
\det \left( {\begin{array}{*{20}c}
   {b_q } & {c_q }  \\
   { - c_q } & {b_{q + Q} }  \\
\end{array}} \right) = 0,
\ee
that is,
\be
\left( {1 - \frac{1}{2}VK_q^{\left( 0 \right)} } \right)
\left( {1 - \frac{1}{2}VK_{q + Q}^{\left( 0 \right)} } \right) 
+ \left( {\frac{1}{2}VK_q^{\left( u \right)} } \right)^2  = 0.
\ee
We perform the analytic continuuation of $i\Omega_n \rightarrow \omega + i\delta$,
and then expand each quantity with respect to $\omega$ and $q$.
Noting that 
\be
\frac{1}{2}K_Q^{\left( 0 \right)} \left( {i\Omega _n  = 0} \right) 
= \frac{1}{N}\sum\limits_{k \in RBZ} {\frac{1}{{E_{\bf k} }}}  = \frac{1}{V},
\ee
we find, after some algebra and a numerical computation,
\be
\omega \simeq 0.87J q.
\ee
Thus, the spin-wave velocity is $c_{sw}=0.87J$.
This result is inagreement with the known established value of 
$c_{sw}=1.65J$
estimated from the quantum Monte Carlo simulations
and a series expansion.\cite{Beard1998,Kim1998,Singh1989}
However, the value of $c_{sw}$ depends on $\chi$.
Since the estimation of $\chi$ based on a perturbative analysis,
the value of $\chi$ would change by further including higher order terms.
Here we do not attempt to estimate $\chi$ precisely in this way 
because it is hard to fix the value from that procedure.
Instead, we compute $\chi$ dependence of $c_{sw}$ by
fixing $\Delta_{st} = 0.60J$, which is the constraint from
the experiments and the numerical simulations.
In Fig.\ref{fig_chi_vs_csw}, $\chi$ versus $c_{sw}$ is shown.
From the known value of $c_{sw}$, $\chi$ is evaluated instead.
$c_{sw}\simeq 1.65$ is obtained by setting $\chi=1.25$.
This value implies that the quasiparticle band width is $1.76J$,
which is a reasonable value compared to the experiment.

\begin{figure}
   \begin{center}
    \includegraphics[width=3.4in]{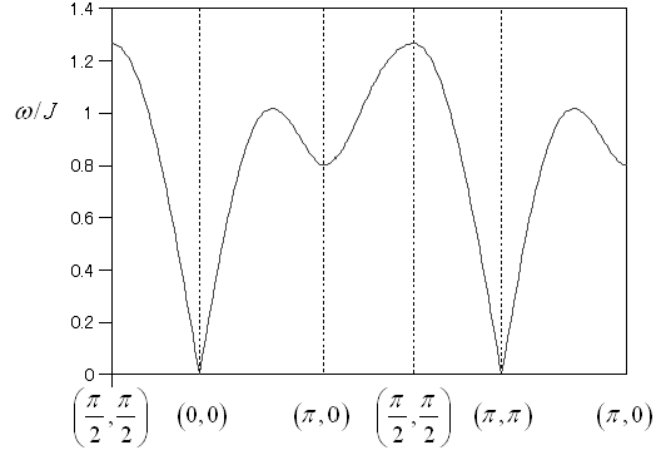}
   \end{center}
   \caption{ \label{fig_spin_wave_disp}
	Numerically calculated spin wave dispersion along high-symmetry directions.
	The lattice size is $40 \times 40$ and $\delta = 0.2$.
    }
 \end{figure}

\begin{figure}
   \begin{center}
    \includegraphics[width=3.4in]{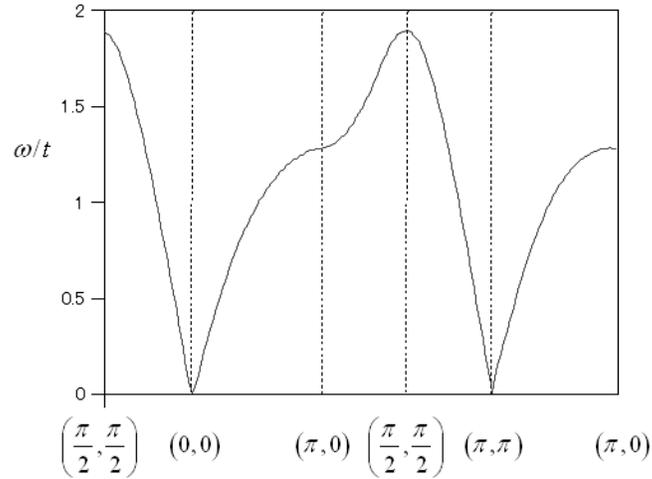}
   \end{center}
   \caption{ \label{fig_sdw_spin_wave_disp}
	Numerically calculated spin wave dispersion along high-symmetry directions 
	for the spin-density wave state at half-filling.
	The lattice size is $40 \times 40$ and $\delta = 0.1$.
	The on-site Coulomb repulsion is $U=5t$.
    }
 \end{figure}

In the limit of $\Delta_{st}/(\chi J) \gg 1$, the spin wave dispersion
is easily obtained\cite{Hsu1990} 
as in the spin-density wave state.\cite{Schrieffer1989,Singh1990}
Using the approximation like
$$
\frac{1}{{E_{\bf k} }} \simeq \frac{1}{{\Delta _{st} }} 
- \frac{{\left| {\kappa _{\bf k} } \right|^2 }}{{2\Delta _{st}^3 }},
$$
and noting 
\be
\frac{1}{N}\sum\limits_{k \in RBZ} {\kappa _{{\bf k} + {\bf q}}^* \kappa _{\bf k} }  
= \frac{{\left( {\chi J} \right)^2 }}{4}\left( {\cos q_x  + \cos q_y } \right),
\ee
we find the following dispersion,
\be
\omega _q  = \frac{{\left( {\chi J} \right)^2 }}{{\Delta _{st} }}
\sqrt {1 - \frac{1}{4}\left( {\cos q_x  + \cos q_y } \right)^2 }.
\label{eq_spin_wave_disp}
\ee
This coincides with the result of the spin wave theory
of the antiferromagnetic Heisenberg model.
The same form of the dispersion is also obtained in the spin-density wave state
in the strong coupling limit.\cite{Schrieffer1989,Singh1990}
Although the assumption of $\Delta_{st}/(\chi J) \gg 1$ is not 
valid for excitations with $\omega > \Delta_{st}$,
eq.(\ref{eq_spin_wave_disp}) is better for high-energy excitations
than numerically obtained spin wave dispersion\cite{Hsu1990}
shown in Fig.\ref{fig_spin_wave_disp} along high-symmetry directions 
in the Brillouin zone.
The situation is similar to the spin density wave state
as shown in Fig.\ref{fig_sdw_spin_wave_disp}.

The full spin wave excitation spectrum is investigated
by the neutron scattering experiments in the undoped compound.\cite{Coldea2001}
Most of features is in good agreement with the spin wave dispersion 
(\ref{eq_spin_wave_disp}) with a suitable prefactor 
except around $(\pi,0)$.
The dip around this point can be explained by including 
a ring exchange term to the Heisenberg model.\cite{Coldea2001,Katanin2002}
Effect of righ-exchange interaction on the $\pi$-flux state is
considered in ref.\citenum{Chung2003} within the mean field theory.
Although it would be interesting to investigate fluctuations about 
the mean field state including the ring-exchange interaction,
consideration of such an effect is beyond the scope of this paper.

\section{Summary and Discussion}
\label{sec_summary}
In this paper, the effect of fluctuations about the $\pi$-flux mean field state
has been investigated.
As for the fluctuations of the Lagrange multiplier field,
it is shown that the fluctuations lead to
a logarithmic potential term in the presence of the mass term.
The mass value is evaluated by solving the 
Schwinger-Dyson equation approximately,
and a reasonable value is obtained.
Because of the mass term,
there is an energy gap in particle-hole excitations.
As a result, the interaction mediated by
the Lagrange multiplier field fluctuations
is long-ranged.
If the fermions were massless, the interaction
would be short-ranged.
(For a different approach to the constraint, 
see ref.\citenum{Dillenschneider2006}.)

As for the fluctuations of the $\pi$-flux state mean fields, 
the lowest order self-energy correction 
doubles the mean field value.
The resulting quasiparticle band withd is much closer to that evaluated 
in the ARPES experiment \cite{Wells1995} compared to the mean field result.
From the pole of the correlation function,
it is found that the amplitude fluctuation mode
has a gap of $1.2J$.
Meanwhile, the phase fluctuations are associated with the spin wave excitation,
which is analyzed by approximating the potential term mediated by 
the Lagrange multiplier field fluctuations by a short-range repulsion.

The dynamical mass generation is consistent with the confinement of fermionic excitations.
That is, there are no low-lying fermionic excitations in the two-dimensional 
Heisenberg antiferromagnet.
Due to the mass term of the fermions on the order of $J$,
the propagation of the Lagrange multiplier field do not excite particle-hole pairs.
Therefore, there is no damping for the propagation.
As a result, there is a long-range interaction between the fermions.
Since the Lagrange multiplier field lives on three spatial and time dimensions,
the interaction between fermions mediated by the exchange of the Lagrange multiplier
fields is a logarithmic potential.
Under the effect of such a confining potential, 
the fermions are confined.
This is, of course, consistent with the constraint.
Because the constraint is satisfied for low-energy excitations.
Contrary, if the fermions were massless, the propagation of
the Lagrange multiplier fields would excite many particle-hole pairs.
The interaction between fermions mediated by the exchange of such a field
is short range interaction.
Therefore, there would be no confinement, or fermionic excitations 
would appear in the low-lying excitations,
and the constraint would be no longer satisfied.
This picture suggests that a small mass term in the Lagrange multiplier field
should break the confinement of the fermions.
This contradicts with the conclusion in ref.\citenum{Pereg-Barnea2003}
in which it is argued that the dynamical mass generation occurs 
in the presence of a small gauge field mass.


As for the application to the high-temperature superconductors,
the theory suggests that the low-lying excitations in the spin disordered regime
is described by the fermions with the background of
the $\pi$-flux state correlations.\cite{Lee1996}
Because in the spin disordered regime there is no staggered magnetization.
Therefore, there is no mass gap for particle-hole excitations.
In this case  the interaction between the fermions mediated by the Lagrange multiplier
fields is of short range interaction.
We may neglect the effect of such a short-range interaction.
Meanwhile, the phase fluctuations are important for the properties of the fermions.
Because the phase fluctuations play the role of transverse gauge field
living in three spatial and time dimensions.
As discussed in ref.\citenum{LeeNagaosa1992},
such the transverse gauge field fluctuations lead to non-Fermi liquid behavior.

In the continuum theory, or the QED$_3$ theory, the gauge field
associated with fluctuations about the mean field state is
either compact or non-compact.
By contrast, in our theory the phase fluctuations are compact 
but the Lagrange multiplier fields are not compact.
This may require a different approach to instanton effects
discussed in the confinement phenomenon in the compact QED$_3$ theory.\cite{Polyakov1977}

\section*{Acknowledgment}
I would like to thank Prof. T. Tohyama for useful discussion.
The numerical calculations were carried out in part 
on Altix3700 BX2 at YITP in Kyoto University.

\appendix
\section{Derivation of the Effective Interaction}
\begin{figure}
   \begin{center}
    \includegraphics[width=2in]{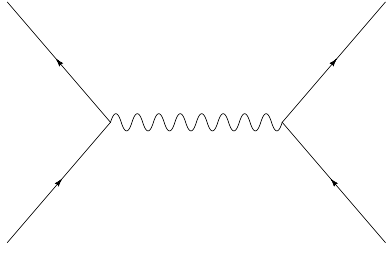}
   \end{center}
   \caption{ \label{fig_zero_th_order}
	The lowest order contribution to the fermion interaction due to
	the exchange of the $\phi_q$ field. Wavy line represents the propagator
	of the $\phi_q$ field and straight lines represent the fermion propagator.
    }	
\end{figure}

\begin{figure}
   \begin{center}
    \includegraphics[width=5in]{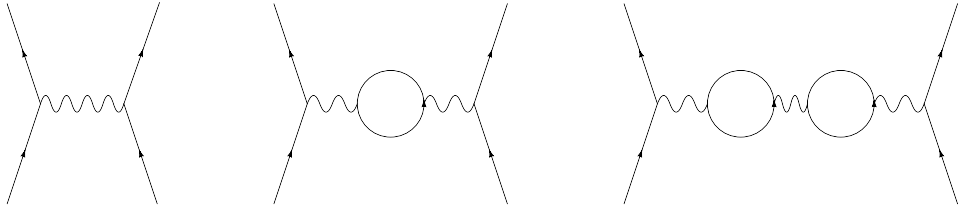}
   \end{center}
   \caption{ \label{fig_higher_orders}
	Feynman diagram representing the random phase approximation.
    }	
\end{figure}

In deriving the effective interaction between the fermions, eq.(\ref{eq_Seff}),
we first integrate over the fermion fields to obtain the effective action
for the auxiliary fields,
and then the effective interaction between the fermions 
is derived using the resulting effective action for the auxiliary fields.
In this Appendix, we show that this kind of calculation is equivalent
to the random phase approximation.
To make clear the point, we consider a following simple model,
\be
H = \sum\limits_k {\xi _k f_k^\dag  f_k }  
+ \sum\limits_{k,q} {\phi _q f_{k + q}^\dag  f_k }  
+ \sum\limits_q {K_q \phi _q \phi _{ - q} }.
\label{eq_appH}
\ee
The lowest order term for the interaction between fermions mediated by
the $\phi_q$ field is given by
\be
V_0  =  - \frac{1}{2}\sum\limits_{k,k',q} 
{\frac{1}{{K_q }}f_{k + q}^\dag  f_{k'}^\dag  f_{k' + q} f_k }.
\ee
The Feynman diagram for this term is represented by Fig.\ref{fig_zero_th_order}.
By including fermion bubble diagrams and summing over terms represented by
Fig.\ref{fig_higher_orders}, we obtain 
\be
V_{RPA}  =  - \frac{1}{2}\sum\limits_{k,k',q} 
{\frac{1}{{2K_q  + \Pi _q \left( {i\Omega _n } \right)}}f_{k + q}^\dag  f_{k'}^\dag  
f_{k' + q} f_k },
\label{eq_V_RPA}
\ee
where
\be
\Pi _q \left( {i\Omega _n } \right) =  
- \frac{1}{{\beta N}}\sum\limits_{i\omega _n ,k} 
{G_k^{\left( 0 \right)} \left( {i\omega _n } \right)
G_{k + q}^{\left( 0 \right)} \left( {i\omega _n  + i\Omega _n } \right)},
\ee
with $G_k(i\omega_n)=1/(i\omega_n-\xi_k)$.

Now we show that the same interaction term is obtained by the following 
path-integral calculation.
We consider the action for the Hamiltonian (\ref{eq_appH}),
\bea
S &=& \sum\limits_k {f_k^\dag  \left( {i\omega _n } \right)
\left( { - i\omega _n  + \xi _k } \right)f_k \left( {i\omega _n } \right)}  
+ \sum\limits_{k,q} {\phi _q \left( {i\Omega _n } \right)f_{k + q}^\dag  
\left( {i\omega _n  + i\Omega _n } \right)f_k \left( {i\omega _n } \right)}  
\nonumber \\
& &
+ \sum\limits_q {K_q \phi _q \left( {i\Omega _n } \right)\phi _{ - q} 
\left( { - i\Omega _n } \right)}.
\eea
Integrating out fermion fields, we obtain
\bea
S_\phi  &=&  - Tr\ln \left[ {\left( { - i\omega _n  + \xi _k } \right)
\delta _{k,k'} \delta _{n,n'}  + \phi _{k - k'} 
\left( {i\omega _n  - i\omega _{n'} } \right)} \right] 
+ \sum\limits_q {K_q \phi _q 
\left( {i\Omega _n } \right)\phi _{ - q} \left( { - i\Omega _n } \right)} \nonumber \\
&=& \sum\limits_q \left[ {K_q  
+ \frac{1}{2}\Pi _q \left( {i\Omega _n } \right)} \right]
\phi _q \left( {i\Omega _n } \right)\phi _{ - q} 
\left( { - i\Omega _n } \right) + ....
\eea
Using the effective action for the $\phi_q$ field obtained this way, 
we consider the following action,
\bea
S' &=& \sum\limits_k {f_k^\dag  \left( {i\omega _n } \right)
\left( { - i\omega _n  + \xi _k } \right)f_k \left( {i\omega _n } \right)}  
+ \sum\limits_{k,q} {\phi _q \left( {i\Omega _n } \right)f_{k + q}^\dag  
\left( {i\omega _n  + i\Omega _n } \right)f_k \left( {i\omega _n } \right)} \nonumber \\
& & + \sum\limits_q \left[ {K_q  + \frac{1}{2}\Pi _q \left( {i\Omega _n } \right)} 
\right]\phi _q \left( {i\Omega _n } \right)\phi _{ - q} 
\left( { - i\Omega _n } \right).
\eea
By integrating out the $\phi_q$ field, we obtain (\ref{eq_V_RPA}).


\end{document}